\documentclass[11pt]{article}
\newlength{\vshift}
\newlength{\hshift}
\usepackage{amssymb,amsopn}

\def\ds{\stackrel{\star}{,}}
\def\x{\hat x}

\def\p{\partial}

\def\lb{\lbrack}
\def\rb{\rbrack}
\def\p{\partial}

\def\bb{\begin{equation}}
\def\eb{\end{equation}}
\def\bea{\begin{eqnarray}}
\def\eea{\end{eqnarray}}

\begin{document}

\begin{titlepage}
%\rightline{LMU-ASC 32/05}

\begin{center}

%\vskip 5em

{\Large{\bf Differential calculus and gauge transformations on a deformed space}}

\vskip 4em

{{\bf Julius Wess${}^{1,2,3}$}}

\vskip 1em
%Lecture given at\\
%{\it Second Meeting on the Interface of Gravitation and Quantum...}\\
%5-8 December, 2005 Zacatecas, Mexico\\

Dedicated to the 60th birthday of Prof. Obregon\footnote{To appear in General Relativity and Gravitation Journal, Obregon's Festschrift 2006}

\vskip 2em

${}^{1}$Arnold Sommerfeld Center for Theoretical Physics\\
Universit\"at M\"unchen, Fakult\"at f\"ur Physik\\
Theresienstr.\ 37, 80333 M\"unchen, Germany\\[1em]

${}^{2}$Max-Planck-Institut f\"ur Physik\\
        F\"ohringer Ring 6, 80805 M\"unchen, Germany\\[1em]

${}^{3}$Universit\"at Hamburg, II Institut f\"ur Theoretische Physik\\
and DESY,Hamburg\\
        Luruper Chaussee 149, 22761 Hamburg, Germany\\[1em]

\end{center}

\vspace{3em}

\abstract{
Deformed gauge transformations on deformed coordinate spaces are considered for any Lie algebra. The representation theory of this gauge group forces us to work in a deformed Lie algebra as well. This deformation rests on a twisted Hopf algebra, thus we can represent a twisted Hopf algebra on deformed spaces. That leads to the construction of Lagrangian invariant under a twisted Lie algebra.
}

\vspace{3em}
%\begin{center}
\noindent This article is based on common work with Paolo Aschieri, Christian Blohmann, Marija Dimitrijevi\' c, Branislav Jur\v co, Frank Meyer, Lutz M\"oller, Stefan Schraml and Peter Schupp.
%\end{center}

%\vspace*{1.5cm}
\vspace{2em}



%\vspace*{1cm}
\vspace{2em}
\quad\scriptsize{E-mail: wess@theorie.physik.uni-muenchen.de}
\vfill

\end{titlepage}\vskip.2cm

\setcounter{page}{1}
\newcommand{\Section}[1]{\setcounter{equation}{0}\section{#1}}
\renewcommand{\theequation}{\arabic{section}.\arabic{equation}}

\Section{Introduction}

Since Newton the concept of space and time has gone through various
changes. All stages, however, had in common the notion of a continuous
linear space. Today we formulate fundamental laws of physics, field theories, gauge
field theories and the theory of  gravity on differentiable manifolds. That
a change in the concept of space for very short distances might be necessary
was already anticipated  in 1854 by Riemann in his famous inaugural lecture \cite{Riemann}

\vspace{0.5cm}
"Now it seems that the empirical notions on which the metric determinations of Space are based, the concept of a solid body and a light ray, lose their validity in the infinitely small; it is therefore quite definitely conceivable that the metric relations of Space in the infinitely small do not conform to the hypotheses of geometry; and in fact, one  ought to assume this as soon as it permits a simpler way of explaining phenomena...

....... An answer to these questions can be found only by starting from that conception of phenomena which has hitherto been approved by experience, for which Newton laid the foundation, and gradually modifying it under the compulsion of facts which cannot be explained by it. Investigations like the one just made, which begin from general concepts, can serve only to ensure that this work is not hindered by too restricted concepts, and that the progress in comprehending the connection of things is not obstructed by traditional prejudices.".

\vspace*{0.5cm}
There are indications today that at very short distances we might have to go beyond differential manifolds.

In contrast to coordinate space, phase space - the space of coordinates and momenta - has seen a more dramatic change. Forced by quantum mechanics
we understand it as an algebraic entity based on Heisenberg's commutation relations for canonical variables
\begin{eqnarray}
\lb x^i, p^j\rb \hspace*{-2mm}&=&\hspace*{-2mm} i\hbar \delta^{ij},\nonumber\\
\lb x^i, x^j \rb = 0, \hspace*{-2mm}&&\hspace*{-2mm} \lb p^i, p^j\rb = 0. \label{i1}
\end{eqnarray}
Space and momenta have become noncommutative, they form an algebra.

This algebraic setting has proved to be extremely successful. We would not understand fundamental facts of physics, the uncertainty relation or the existence of atoms, e.g., without it.

The uncertainty relation, however, brings us in conflict with Einstein's law of gravity if we assume continuity in the space variable for arbitrary small distances \cite{2}. From the uncertainty relation
\begin{equation}
\Delta x^i \cdot \Delta p^j \ge \frac{\hbar}{2}\delta^{ij}, \label{i2}
\end{equation}
follows that we need very high energies to measure very short distances.
High energies lead to the formation of black holes with a Schwarzschild radius
proportional to the energy. In turn, this does not allow
the measurement of distances smaller than the Schwarzschild radius.

This is only one of several arguments that we have to expect some
changes in physics for very small distances. Other arguments are based
on the singularity problem in Quantum field theory and the fact that
Einstein's theory of gravity is non-renormalizable when quantized \cite{2}.

Why not try an algebraic concept of space time that could guide us to changes in our present formulation of laws of physics? This is different from the discovery of quantum mechanics. There physics data forced us to introduce the concept of noncommutativity. Now we take noncommutativity as a guide into an area of physics where physical data are almost impossible to obtain. We hope that it might solve some conceptual problems that are still left at very
small distances. We also hope that it could lead to predictions that can be tested in not too far a future by experiment.

The idea of noncommutative coordinates is almost as old as quantum field
theory. Heisenberg proposed it in a letter to Peierls \cite{heisen} to solve the
problem of divergent integrals in relativistic quantum field theory. The
idea propagated via Pauli to Oppenheimer. Finally H.~S.~Snyder, a student
of Oppenheimer, published the first systematic analysis of a quantum theory
built on noncommutative spaces \cite{snyder}. Pauli called this work mathematically
ingenious but rejected it for reasons of physics \cite{5}.

In the meantime the theory of renormalization has found a reasonable answer
to the divergency problem in quantum field theory. We should not forget,
however, that it was the renormalization problem that led to quantum gauge theories and to supersymmetric theories. Only Einstein's theory of gravity remained unrenormalizable when quantized.

From quantum spaces and quantum groups new mathematical concepts have
emerged by the pioneering work of V.~G.~Drinfel'd, L.~Faddeev, M.~Jimbo and I.~Manin
\cite{drinfeld}. This also revived the interest in noncommutativity in physics.

Flatto and Sternheimer \cite{flato} have developed the machinery of deformation quantization. There noncommutativity appears in the form
of noncommutative products of functions of commutative variables. These
products are called star products ($\star$-products). They deform the
commutative algebras of functions based on pointwise multiplication to noncommutative algebras based on the star product.

Deformation theory has reached a very high and powerful level by the
work of Kontsevich and his formality theorem \cite{kontsevich}.

These developments make it worthwhile to reexamine the concept of noncommutative coordinates in physics. 
We first show that the points of view of noncommutative coordinates and of noncommutative $\star$-products 
are intimately related.

\Section{The algebra}

It is the algebraic structure\footnote{Algebra will always refer to associative algebras.} of continuous spaces that we want to deform.
To show this structure we first consider polynomials in commutative variables
$x^1,\dots, x^N$ with complex coefficients. To define them we first define the algebra
over $\mathbb C$,  freely generated by the variables $x^1,\dots, x^N$
\begin{equation}
{\cal A} = \mathbb{C}[x^1, \dots, x^N] .\label{a1}
\end{equation}
This means that we take all the formal products of the $N$ elements $x^1,\dots, x^N$ as
a basis for a linear space over $\mathbb{C}$. A different ordering in the coordinates gives rise to an independent element of the basis! Multiplication  of the basis elements is natural. This then defines the freely generated algebra.

Next we consider the relations
\begin{equation}
{\cal R}_x: \quad\quad [x^i, x^j] = 0. \label{a2}
\end{equation}
They generate an ideal (left and right). The quotient
\begin{equation}
{\cal P}_x = \frac{{\cal A}}{I_{{\cal R}}} \label{a3}
\end{equation}
is the algebra of polynomials in $N$ commuting variables. The definition
of the algebra ${\cal P}_x$ can be extended by including formal series in the basis
elements. This is denoted by two brackets
\begin{equation}
{\cal A}_x = \frac{\mathbb{C}[[x^1, \dots, x^N]]}{I_{{\cal R}}}. \label{a4}
\end{equation}

Up to now we have used algebraic concepts only. No topological properties have been mentioned. Our ambition is to go as far as possible in developing a deformed differential calculus without invoking topological properties.

A natural way is to deform the relation (\ref{a2}). The $N $variables we shall now dress with a hat
\begin{equation}
\hat{{\cal R}}:\quad\quad [\hat{x}^i, \hat{x}^j] - ihC^{ij}(\hat{x}) = 0, \label{a5}
\end{equation}
where $h$ is a deformation parameter. For $h=0$ we obtain the usual algebra
of commuting variables as introduced above. 

The relations again generate a both-sided ideal $I_{\hat{{\cal R}}}$
\begin{equation}
I_{\hat{{\cal R}}}: \quad\quad (\hat{x}\dots\hat{x})\Big( [\hat{x}^i, \hat{x}^j] - ihC^{ij}(\hat{x})\Big) (\hat{x}\dots\hat{x}), \label{a6}
\end{equation}
where $(\hat{x}\dots\hat{x})$ stands for an arbitrary product of $\hat{x}$ in the freely generated algebra ${\hat{\cal A}}$. Multiplying an element of $I_{\hat{{\cal R}}}$ by an element of ${\hat{\cal A}}$ from the right or left yields an element of $I_{\hat{{\cal R}}}$ again. The quotient
\begin{equation}
\hat{{\cal A}}_{\hat{x}} =
\frac{\mathbb{C}[[\hat{x}^1, \dots, \hat{x}^N]]}{I_{{\hat{\cal R}}}}. \label{a7}
\end{equation}
is an  algebra in the noncommuting coordinates $\hat{x}$.

The art of the game now is to find reasonable relations. They of course should
be algebraically consistent. This restricts the $\hat{x}$ dependence of $C^{ij}(\hat{x})$. It has to be antisymmetric in $i$ and $j$. Less trivial is the Jacoby identity which restricts the possible choices of $C^{ij}(\hat{x})$ very much.

To be consistent with the reality property $(x^i)^* = x^i$ we demand a conjugation for $\hat{x}$ as well
\begin{equation}
(\hat{x}^i)^* = \hat{x}^i, \quad
(\hat{x}^i\hat{x}^j)^* = (\hat{x}^j)^* (\hat{x}^i)^*, \quad (i)^* = -i. \label{a8}
\end{equation}
This implies
\begin{equation}
(C^{ij})^* = -C^{ji} = C^{ij}. \label{a9}
\end{equation}

Well-known examples of such algebras are:

\begin{enumerate}
\item
The deformation with $\hat{x}$-independent constant $C^{ij}$. This is the same algebra in coordinate space as the Heisenberg algebra in phase space. We will call it the canonical or for short $\theta$-deformation
\begin{equation}
[\hat{x}^i, \hat{x}^j] = ih\theta^{ij}. \label{a10}
\end{equation}

\item
The Lie algebra type of deformation. In this case $C^{ij}(\hat{x})$ is linear in $\hat{x}$
\begin{equation}
[\hat{x}^i, \hat{x}^j] = ihf^{ijk}\hat{x}^k. \label{a11}
\end{equation}
with real structure constants $f^{ijk}$. The algebra $\hat{{\cal A}}_{\hat{x}}$ that we are constructing is the enveloping algebra of the Lie algebra (\ref{a11}).

A particularly interesting example of the Lie algebra type is
\begin{equation}
[ \x^\mu , \x^\nu ] = i (a^\mu\x^\nu - a^\nu\x^\mu) \label{a12}
\end{equation}
with real parameters $a^\mu$. In a basis where $a^i=0$ for $i\neq N$ and $a^N=1/\kappa$ we can identify this algebra with the algebra of the $\kappa$-deformations \cite{luk}.

\end{enumerate}

The size of the algebra $\hat{{\cal A}}_{\hat{x}}$ will depend on the ideal $I_{{\hat{\cal R}}}$. It can range from empty to the freely generated algebra itself. We certainly would like an infinite algebra, if possible of the "size" of the algebra ${\cal A}_x$ of
commuting variables. To be more precise, the vector space of the algebra
${\cal A}_x$  can be decomposed into subspaces $V_r$ spanned by monomials of degree $r$.
These vector spaces are finite-dimensional. For the analoguous spaces $\hat{V}_r$ we demand the same dimension. Thus $V_r$ and $\hat{V}_r$ will be isomorphic.

This property is called the Poincar\' e-Birkhoff-Witt property. The $\theta$-deformation and the enveloping Lie algebras have this property.

\Section{The star product}

In this section we extend the vector space isomorphism
\begin{equation}
\hat{V}_r \sim V_r \label{sp1}
\end{equation}
to an  algebra morphism
\begin{equation}
\hat{{\cal A}}_{\hat{x}} \sim {\cal A}_x .\label{sp2}
\end{equation}
Schematically we proceed as follows: We first choose a basis in
$V_r$ and $\hat{V}_r$. The fully symmetrized monomials are a natural
choice but not the only one. By the vector space ismorphism we map
polynomials
\begin{equation}
p_l(x) \longleftrightarrow \hat{p}_l(\hat{x}), \quad l=1, 2 \label{sp3}
\end{equation}
by a map of the basis. Polynomials can be multiplied
\begin{equation}
\hat{p}_1(\hat{x})\cdot \hat{p}_2(\hat{x}) =
\widehat{p_1p_2}(\hat{x}) .\label{sp4}
\end{equation}
By the isomorphism of (\ref{sp3}) we map this polynomial back to a polynomial in ${\cal A}_x$
\begin{equation}
\widehat{p_1p_2}(\hat{x}) \mapsto p_1(x)\star p_2(x) .\label{sp5}
\end{equation}
This defines the star product of two functions. It is bilinear and associative
but noncommutative.

For the $\theta$-deformation in the symmetric basis we obtain \cite{weyl}
\begin{equation}
p_1(x)\star p_2(x) = \mu \Big(
e^{\frac{i}{2}h\theta^{\rho\sigma}\p_\rho
\otimes\p_\sigma}p_1(x)\otimes p_2(x)\Big)  , \label{sp6}
\end{equation}
where $\mu$ is the multiplication map
\begin{equation}
\mu \left( f(x)\otimes g(x) \right) = f(x)\cdot g(x). \label{sp7}
\end{equation}
This $\star$-product is the well-known Moyal-Weyl product. It can be extended to $C^\infty$ functions, remaining bilinear and associative. The power series in $h$ will not converge for arbitrary $C^\infty$ functions, we in general consider it as a formal power series.

When we expand in $h$ we obtain
\begin{eqnarray}
f(x)\star g(x)\Big| _{h=0} = f(x)g(x) \nonumber
\end{eqnarray}
and
\begin{eqnarray}
\frac{1}{h}\Big( f(x)\star g(x) - g(x)\star f(x)\Big) \Bigg| _{h=0}
= \frac{i}{2}\theta^{\rho\sigma} \Big( (\partial_\rho
f(x))(\partial_\sigma g(x)) - (\partial_\rho g(x))(\partial_\sigma
f(x)) \Big) .\nonumber
\end{eqnarray}
This is a Poisson structure.

Kontsevich has shown that for any Poisson structure on a differential manifold there exists a $\star$-product deformation.
Knowing this, it seems natural to investigate noncommutative spaces in the
$\star$-product framework. 

Our aim now is to formulate laws of physics on an algebra of functions whose product is not the pointwise porduct but a noncommutative star product. We call this algebra ${\cal A}^\star {\cal F}_x$.

One important step in this direction is the development of a differential
calculus on this deformed algebra of functions ${\cal A}^\star {\cal F}_x$. This we will do next. But let me for the convenience of the reader summarize the notation first.

\vspace*{5mm}

{\bf Notation}

\begin{itemize}

\item
$\mu (f\otimes g) = f\cdot g$ - pointwise multiplication,

\item
%\vspace*{-1cm}
\begin{eqnarray}
\mu^\star (f\otimes g) \hspace*{-2mm}&=&\hspace*{-2mm} f\star g =
\mu \Big( e^{\frac{i}{2}h\theta^{\rho\sigma}\p_\rho  \otimes\p_\sigma} f\otimes g\Big) \nonumber\\
&=&\hspace*{-2mm} \sum _{n=0}^\infty \Big(\frac{ih}{2}\Big)^n
\frac{1}{n!}\theta^{\rho_1\sigma_1}\dots
\theta^{\rho_n\sigma_n}\Big(\partial_{\rho_1}\dots\partial_{\rho_n}f\Big)
\Big( \partial_{\sigma_1}\dots\partial_{\sigma_n} g\Big) \nonumber\\
&& \nonumber
 {\mbox{   - star multiplication,}} \nonumber
\end{eqnarray}

\item
${\cal A}$ - freely generated algebra of $N$ independent variables,

\item
${\cal P}_x$ - algebra of polynomials in $N$ commuting varables ${x_1, \dots, x_N}$,

\item
$\hat{\cal{P}}_{\hat{x}}$ - algebra of polynomials in $N$ noncommuting variables ${\hat{x}_1, \dots, \hat{x}_N}$,

\item
${\cal A}_x$ - algebra of formal power series of polynomials in $N$ commting variables,

\item
$\hat{\cal{A}}_{\hat{x}}$ - algebra of formal power series of polynomials in $N$ noncommuting variables,

\item
$V_r$ - linear subspace of ${\cal A}_x$ spanned by monomials in ${x_1, \dots, x_N}$ of degree $r$,

\item
$\hat{V}_r$ - linear subspace of $\hat{\cal{A}}_{\hat{x}}$ spanned by monomials in ${\hat{x}_1, \dots, \hat{x}_N}$ of degree $r$,

\item
${\cal F}_x$ - linear space of functions in $N$ commuting variables ${x_1, \dots, x_N}$,

\item
${\cal A}{\cal F}_x$ - algebra of functions in $N$ commuting variables ${x_1, \dots, x_N}$ with pointwise multiplication,

\item
${\cal A}^\star{\cal F}_x$ - algebra of functions in $N$ commuting variables ${x_1, \dots, x_N}$ with $\star$-product multiplication,

\item
${\cal D}_{ \{d\} }$ - higher order differential operator acting on ${\cal A}{\cal F}_x$

\item
${\cal D}^\star_{ \{d\} }$ -  higher order differential operator acting on  $\hat{{\cal A}}{\cal F}_x$,

\item
${\cal A}{\cal D}_{ \{d\} }$ - algebra of higher order differential operators $D_{ \{d\} }$,

\item
${\cal A}^\star {\cal D}^\star_{ \{d\} }$ - algebra of higher order differential operators $D^\star_{ \{d\} }$.

\end{itemize}

\Section{A deformed differential calculus}

Let us first define a derivative as a map of $C^\infty$ functions to $C^\infty$ functions
\begin{eqnarray}
\partial_\mu : & & \quad {\cal F}_x \to {\cal F}_x \nonumber\\
f (x) &\mapsto & (\partial_\mu f (x)) .\label{d1}
\end{eqnarray}
For polynomials this map can be defined purely algebraically by stating the rule
\begin{equation}
\partial_\mu:\quad\quad x^\rho\mapsto \delta^\rho_\mu \label{d2}
\end{equation}
and using the Leibniz rule
\begin{equation}
(\p_\mu (p_1\cdot p_2)) = (\p_\mu p_1)\cdot p_2 + p_1\cdot (\p_\mu p_2) .\label{d3}
\end{equation}
We know that this defines the derivative of polynomials and their
products. This can be easily extended to formal power series. We use it to define the derivative on ${\cal A}^\star{\cal F}_x$ by first mapping an element of ${\cal A}^\star{\cal F}_x$ to ${\cal A}{\cal F}_x$,
differentiate this element in ${\cal A}{\cal F}_x$ and map it back to ${\cal A}^\star{\cal F}_x$. Thus, we define
\begin{eqnarray}
\hat{\partial}_\mu: & & \quad {\cal A}^\star{\cal F}_x \to {\cal A}^\star{\cal F}_x \nonumber\\
\hat{f}(\hat{x}) &\mapsto & (\partial_\mu f (x)) \mapsto
(\hat{\partial}_\mu \hat{f} (\hat{x})).\label{d4}
\end{eqnarray}
The basis  can always be arranged such that (\ref{d2}) remains
\begin{equation}
\hat{\partial}_\mu:\quad\quad \hat{x}^\rho\mapsto \delta^\rho_\mu, \label{d5}
\end{equation}
but the Leibniz rule will change in general. We derive it in the star product formalism and use the fact that $f\star g$ is a function again
\begin{equation}
\p_\mu (f\star g) = (\p_\mu f)\star g + f\star (\p_\mu g) + f(\p_\mu\star) g .\label{d6}
\end{equation}

In the case of $\theta$-deformation the $\star$-operation is
$x$-independent and we obtain the usual Leibniz rule:
\begin{equation}
\p_\mu (f\star g) = (\p_\mu f)\star g + f\star (\p_\mu g) .\label{d7}
\end{equation}
%()-deformation:
For $x$-dependent $\star$-products-the $\kappa$-deformation e.g.-this will
change.

With the help of the Leibniz rule we can extend the concept of derivatives acting on functions to the concept of differential operators
\begin{equation}
\p^\star_\mu f\star  = (\p^\star_\mu f)\star + f\star \p^\star_\mu ,\quad {f\in {\cal A}^\star{\cal F}_x}. \label{d8}
\end{equation}

We have obtained a differential calculus on the deformed algebra of
functions ${\cal A}^\star{\cal F}_x$. When the derivative acts on a
deformed algebra of functions we sahll emphasize this by a star on
the derivative.

This calculus we can extend to higher order differential operators.

On the commutative algebra ${\cal A}{\cal F}_x$ a higher order differential operator is defined as follows
\begin{equation}
D_{\{d\}} = \sum_{r\ge 0}d_r^{i_1\dots i_r}(x)\frac{\p}{\p x^{i_1}}\dots\frac{\p}{\p x^{i_r}} .\label{d9}
\end{equation}
These operators form an algebra ${\cal A}D_{ \{d\} }$ because we know how to multiply them.

For the deformed space of functions we define the deformed differential operators
\begin{equation}
D_{\{d\}}^\star = \sum_{r\ge 0}d_r^{i_1\dots i_r}(x)\star\p^\star_{i_1}\star\dots\star\p^\star_{i_r} .\label{d10}
\end{equation}
From the generalized Leibniz rule (\ref{d8}) and the definition of the $\star$-product for functions we learn how to  multiply these deformed operators and in this way obtain the algebra of deformed differential operators ${\cal A}^\star {\cal D}^\star_{ \{d\} }$ acting on elements of the deformed algebra of functions.

There is a formal isomorphism between the algebras ${\cal A}{\cal D}_{ \{d\} }$ and ${\cal A}^\star{\cal D}^\star_{ \{d\} }$. We are going to show this for special subalgebras.

\Section{A map of differential operators}

There exists a higher order differential operator $X^\star_f \in {\cal A}^\star {\cal D}^\star_{ \{d\} }$ such that
\begin{equation}
X^\star_f \star g = f\cdot g \label{5.1}
\end{equation}
where $f$ is an element of ${\cal A}{\cal F}_x$ acting on $g$, an
element of ${\cal F}_x$. To find $X^\star_f$ we proceed as follows
\begin{eqnarray}
f\cdot g \hspace*{-2mm}&=&\hspace*{-2mm} \mu \Big(
e^{\frac{ih}{2}\theta^{ij}\p_i \otimes\p_j}e^{-\frac{ih}{2}\theta^{kl}\p_k \otimes\p_l}
f\otimes g\Big) \nonumber\\
&=&\hspace*{-2mm}\mu \Bigg(
e^{\frac{ih}{2}\theta^{ij}\p_i \otimes\p_j}\sum _{r=0}^\infty \Big(-\frac{ih}{2}\Big)^r
\frac{1}{r!}\theta^{k_1 l_1}\dots
\theta^{k_r l_r}\Big(\partial_{k_1}\dots\partial_{k_r}f\Big) \otimes
\Big( \partial_{l_1}\dots\partial_{l_r} g\Big) \Bigg) \nonumber\\
&=&\hspace*{-2mm}\sum _{r=0}^\infty \Big(-\frac{ih}{2}\Big)^r
\frac{1}{r!}\theta^{k_1 l_1}\dots
\theta^{k_r l_r}\Big(\partial_{k_1}\dots\partial_{k_r}f\Big)\star
\Big( \partial^\star_{l_1}\dots\partial^\star_{l_r} g\Big).\label{gt1}
\end{eqnarray}
The operator we are looking for is
\begin{equation}
X^\star_f = \sum _{r=0}^\infty \Big(-\frac{ih}{2}\Big)^r
\frac{1}{r!}\theta^{k_1 l_1}\dots
\theta^{k_r l_r}\Big(\partial_{k_1}\dots\partial_{k_r}f\Big)\star
\partial^\star_{l_1}\dots\partial^\star_{l_r} .\label{gt2}
\end{equation}
It is a higher order differential operator $\star$-acting on ${\cal
F}_x$.

Because $f\cdot g$ is again an element of ${\cal F}_x$ we can act
with $X^\star_h$ on it
\begin{equation}
h\cdot f\cdot g = (h\cdot f)\cdot g = X^\star_{(hf)}\star g = h\cdot
(f\cdot g) = h (X^\star_f\star g )= X^\star_h \star ( X^\star_f\star
g ). \label{gt3}
\end{equation}
It follows that $X^\star_f$ represent the algebra ${\cal A}{\cal F}_x$
\begin{equation}
X^\star_g \star X^\star _f = X^\star _{gf} .\label{5.5}
\end{equation}

Let us consider vector fields
\begin{equation}
\xi = \xi^\mu (x)\partial_\mu .\label{dif1}
\end{equation}
Their product is again in ${\cal A}{\cal D}_{\{d\}}$
\begin{equation}
\xi\eta = \xi^\mu (x)\big( \partial_\mu\eta^\rho(x)\big) \partial_\rho +
\xi^\mu (x)\eta^\rho(x) \partial_\mu\partial_\rho .\label{dif2}
\end{equation}
Through the Lie bracket the vector fields form an algebra
\begin{eqnarray}
\lb \xi, \eta \rb \hspace*{-2mm}&=&\hspace*{-2mm} \Big( \xi^\mu (\p_\mu \eta^\rho) -\eta^\mu(\p_\mu \xi^\rho) \Big)\partial_\rho \nonumber\\
\hspace*{-2mm}&=&\hspace*{-2mm}(\xi\times \eta)^\rho \partial_\rho =  \xi\times \eta . \label{dif3}
\end{eqnarray}
The vector field $\xi$ can be mapped to ${\cal A}^\star{\cal D}^\star_x$
\begin{eqnarray}
\xi^\star\star f \hspace*{-2mm}&=&\hspace*{-2mm} X^\star_{\xi^\rho}\partial^\star_\rho \star f = X^\star_{\xi^\rho}\star \partial_\rho f \nonumber\\
&=&\hspace*{-2mm}\xi\cdot f .\label{dif5}
\end{eqnarray}
From the associativity in the algebra it follows
\begin{equation}
(\eta^\star \star \xi^\star )\star f = \eta^\star\star(\xi^\star \star f)
= \eta^\star\star(\xi f) = \eta\xi f \label{dif6}
\end{equation}
and therefore
\begin{equation}
\eta^\star \star\xi^\star - \xi^\star\star\eta^\star = (\eta\times\xi)^\star .\label{dif7}
\end{equation}
The deformed vector fields under the deformed Lie bracket form the same algebra as the vector fields under the ordinary Lie bracket.

\Section{Gauge transformations}

Ordinary gauge transformations are Lie algebra valued
\begin{eqnarray}
\alpha (x) \hspace*{-2mm}&=&\hspace*{-2mm} \alpha^a(x) T^a ,\nonumber\\
\lb T^a, T^b \rb \hspace*{-2mm}&=&\hspace*{-2mm} if^{abc}T^c .\label{gt5}
\end{eqnarray}
The action on a field is
\begin{equation}
\delta_\alpha \psi = i\alpha\psi = i\alpha^a(x) T^a\psi . \label{gt6}
\end{equation}
This can be reproduced by a star action on the field:
\begin{equation}
\delta^\star_\alpha \star\psi = iX^\star_{\alpha^a(x)}\star T^a\psi
= i\alpha\cdot\psi ,\label{gt7}
\end{equation}
and represents the algebra via star commutators:
\begin{equation}
\lb \delta^\star_\alpha \ds \delta^\star_\beta \rb 
 = i\delta^\star_{[\alpha ,\beta] }.\label{gt8}
\end{equation}
Gauge transformations of this kind have been introduced in
 \cite{defgt}. Interesting is the transformation law of products of fields.

In the {\it undeformed} case we start from the  transformation properties of the individual fields and transform the product as follows:
\begin{eqnarray}
\delta_\alpha (\psi\chi)  \hspace*{-2mm}&=&\hspace*{-2mm} (\delta_\alpha\psi )\chi + \psi (\delta_\alpha\chi )  \nonumber\\
&=&\hspace*{-2mm} i\alpha^a\Big( (T^a\psi) \chi +\psi(T^a\chi) \Big) 
\label{gt10}
\end{eqnarray}
In accordance with (\ref{gt7}) we translate this to a star action
\begin{equation}
\label{gt12}
\delta_\alpha^\star (\psi\star \chi) = i X_{\alpha^a}^\star \star \left\{
(T^a\psi)\star \chi +\psi \star (T^a\chi )\right\}\,.
\end{equation}
The transformation law (\ref{gt12}) represents the algebra as in (\ref{gt8}).

The star product $\psi\star\chi$ is a function again and (\ref{5.1}) can be applied. We obtain 
\begin{equation}
\label{gt13}
\delta_\alpha^\star (\psi\star \chi) = i {\alpha^a}\cdot \left\{
(T^a\psi)\star \chi +\psi \star (T^a\chi )\right\}\,.
\end{equation}
Expanding the function in the bracket to first order in $\theta$ we obtain
\begin{eqnarray}
\delta^\star_\alpha (\psi\star\chi )  \hspace*{-2mm}&=&\hspace*{-2mm} i\alpha^a\{ T^a\psi\cdot\chi + \psi \cdot T^a \chi \nonumber \\
 &+&\hspace*{-2mm}\frac i2 \theta^{\rho\sigma} (T^a\partial _\rho\psi\cdot \partial_\sigma\chi +\partial _\rho\psi\cdot T^a\partial_\sigma\chi ) +\dots\} . \label{gt14}
\end{eqnarray}
To compare this with the first part of (\ref{gt10}) we introduce the star product again but combined with 
$\delta^\star_\alpha\psi = i\alpha\psi$ and the same for $\chi$. 
\begin{eqnarray}
\delta^\star_\alpha (\psi\star\chi )  \hspace*{-2mm}&=&\hspace*{-2mm} (i\alpha \psi)\star\chi + \psi \star (i\alpha \chi) \nonumber \\
 &-&\hspace*{-2mm}\frac i2 \theta^{\rho\sigma} ( (i\partial_\rho \alpha^a)T^a\psi( \partial_\sigma\chi ) + (\partial _\rho\psi)(i\partial_\sigma \alpha^a) T^a\chi ) +  O(\theta^2)\} . \label{gt15}
\end{eqnarray}
This expression can be extended to all orders in $\theta$ by induction. The result is 
\begin{eqnarray}
\delta^\star_\alpha (\psi\star\chi )  \hspace*{-2mm}&=&\hspace*{-2mm}
i\sum_{n=0}^\infty\frac{1}{n!} (-\frac i2)^n \theta^{\rho_1\sigma_1}\dots \theta^{\rho_n\sigma_n}
\{(\partial_{\rho_1}\dots\partial_{\rho_n}\alpha)\psi\star (\partial_{\sigma_1}\dots\partial_{\sigma_n}\chi) 
\nonumber\\ &+&
(\partial_{\rho_1}\dots\partial_{\rho_n}\psi)\star(\partial_{\sigma_1}\dots\partial_{\sigma_n}\alpha)\chi \} .
\end{eqnarray}

The transformation law of the product of fields follows from the transformation law of the tensor product which in turn is a part of the Hopf algebra structure. For {\it undeformed} gauge transformations we can write (\ref{gt10})  in the Hopf algebra language:
\begin{equation}
\delta_\alpha (\psi\otimes\chi) = i\Delta(\alpha)\psi\otimes\chi
\label{gt16}
\end{equation}
The coproduct $\Delta(\alpha)$ represents the Lie algebra in the tensor product representation 
\begin{eqnarray}
&&\Delta(\alpha ) = \alpha\otimes 1 +1\otimes\alpha,  \\
&&[\Delta(\alpha),\Delta(\beta)] =\Delta([\alpha,\beta]) .
\label{gt17}
\end{eqnarray}

The transformation law of the pointwise product can be defined with the multiplication $\mu$:
\begin{equation}
\delta_\alpha (\psi\chi) = \mu \{ \Delta(\alpha)\psi\otimes\chi \} . \label{gt18}
\end{equation}
The transformation law of the $\star$-product can be defined with the $\star$-multiplication  $\mu^\star$ and a twisted coproduct. We define a twisted coproduct:
\begin{equation}
\Delta_{\cal F}(\alpha) ={\cal F}(\alpha\otimes 1+ 1\otimes\alpha ){\cal F}^{-1}
\end{equation}
with
\begin{equation}
{\cal F}= e^{-\frac i2 \theta^{\mu\sigma}\partial_\mu\otimes\partial_\sigma} . \label{gt19}
\end{equation}
This twist ${\cal F}$ has all the properties that are required to define a twisted Hopf algebra structure. We can show that 
\bb
\delta^\star_\alpha (\psi\star\chi)  = 
\mu^\star\{ \Delta_{\cal F}(\alpha )\psi\otimes \chi\} \label{gt22}
\eb
in the $\theta$-expansion by direct calculation.

\Section{Differmorphism}

The usual algebra of differmorphism is generated by vector fields.
They are elements of ${\cal A}{\cal D}_x$
\begin{equation}
\xi = \xi^\mu (x)\partial_\mu .\label{dif1'}
\end{equation}
Their product in ${\cal A}{\cal D}_x$ is
\begin{equation}
\xi\eta = \xi^\mu (x)\big( \partial_\mu\eta^\rho(x)\big) \partial_\rho +
\xi^\mu (x)\eta^\rho(x) \partial_\mu\partial_\rho .\label{dif2'}
\end{equation}
Through the Lie bracket we obtain the Lie algebra of differmorphism
\begin{eqnarray}
\lb \xi, \eta \rb \hspace*{-2mm}&=&\hspace*{-2mm} \Big( \xi^\mu (\p_\mu \eta^\rho) -\eta^\mu(\p_\mu \xi^\rho) \Big)\partial_\rho \nonumber\\
\hspace*{-2mm}&=&\hspace*{-2mm}(\xi\times \eta)^\rho \partial_\rho =  \xi\times \eta . \label{dif3'}
\end{eqnarray}
The vector field $\xi$, an element of  ${\cal A}{\cal D}_x$, can be mapped to ${\cal A}^\star{\cal D}^\star_x$
\begin{eqnarray}
{\cal A}{\cal D}_x &\to & {\cal A}^\star{\cal D}^\star_x \nonumber\\
\xi &\mapsto & X^\star _{\xi^\rho}\partial^\star_\rho = \xi^\star .\label{dif4'}
\end{eqnarray}
When it $\star$-acts on a function $f\in {\cal A}{\cal F}_x$ we obtain
\begin{eqnarray}
\xi^\star\star f \hspace*{-2mm}&=&\hspace*{-2mm} X^\star_{\xi^\rho}\partial^\star_\rho \star f = X^\star_{\xi^\rho}\star \partial_\rho f \nonumber\\
&=&\hspace*{-2mm}\xi\cdot f .\label{dif5'}
\end{eqnarray}
This is analoguous to (\ref{gt2}) and we can proceed as there. From associativity in
${\cal A}{\cal D}_x$ and ${\cal A}^\star{\cal D}^\star_x$ follows
\begin{equation}
(\eta^\star \star \xi^\star )\star f = \eta^\star\star(\xi^\star \star f)
= \eta^\star\star(\xi f) = \eta\xi f \label{dif6'}
\end{equation}
and therefore
\begin{equation}
\eta^\star \star\xi^\star - \xi^\star\star\eta^\star = (\eta\times\xi)^\star .\label{dif7'}
\end{equation}
The deformed vector fields under the deformed Lie bracket form the same algebra. They represent the deformed algebra of differmorphisms.

The Riemann-Einstein theory of gravity has been constructed on this deformed algebra of differmorphisms \cite{grav}. The coproduct of the diffeomorphism algebra has to be modified as before for gauge theories. It is the first theory of gravity defined on a deformed space and is under  investigation now.

\Section{Conclusion}

The formalism developped here opens a way to construct a deformation of differential geometry and therefore deformed gauge theories and gravity theories. Mathematically it is certainly an interesting possibility. If physics knows anything about it is hard to say. Future investigation might shed some light on this question. I am left to quote Riemann and express my hope that:

\vspace{0.5cm}

"...this work is not hindered by too restricted concepts and that the progress in comprehending the connection of things is not obstructed by tradtional prejudices.".

\end{document}